\def\et{{\it et\thinspace al.\ }}    
\def\deg{\ifmmode^\circ\else$^\circ$\fi}    
\def\solar{\ifmmode_{\mathord\odot}\else$_{\mathord\odot}$\fi}
\def\arcs{\ifmmode {'' }\else $'' $\fi}     
\def\arcm{\ifmmode {' }\else $' $\fi}     
\def\buildrel#1\over#2{\mathrel{\mathop{\null#2}\limits^{#1}}}
\def\arcmper{\ifmmode \rlap.{' }\else $\rlap{.}' $\fi}
\def\degper{\ifmmode \rlap.^{\circ }\else $\rlap{.}^\circ $\fi}
\def\arcsper{\ifmmode \rlap.{'' }\else $\rlap{.}'' $\fi}
\def\refrule{\vrule width 1cm depth 0.0pt height 0.4pt.\ }

\documentstyle[aaspp]{article}
\lefthead{Drukier \et}
\righthead{Star counts in NGC~6397}

\begin{document}

\title{Star counts in NGC~6397\altaffilmark{1}}

\author{G. A. Drukier\altaffilmark{2}, G. G. Fahlman, and  H.B. Richer}
\affil{Dept of Geophysics and Astronomy, University of British Columbia,
Vancouver, B.C., V6T~1Z4}
\affil{Electronic mail: drukier@mail.ast.cam.ac.uk, fahlman@geop.ubc.ca,
richer@geop.ubc.ca}
\authoraddr{Institute of Astronomy, Madingley Rd., Cambridge, CB3 OHA, England}

\and

\author{L. Searle, and I. Thompson}
\affil{The Observatories, Carnegie Institution of Washington, 813 Santa Barbara
St., Pasadena,~CA, 91101-1292}
\affil{Electronic mail: searle@ociw.edu, thompson@ociw.edu}

\altaffiltext{1}{Based on observations obtained at the Las Campanas Observatory
of the Carnegie
Institution of Washington.}
\altaffiltext{2}{Present address: Institute of Astronomy, Madingley Rd.,
Cambridge, CB3 OHA, England}

\begin{abstract}
$I$-band CCD images of a large area of the nearby globular cluster NGC~6397
have been used to construct a surface density profile and two luminosity and
mass functions.  The surface density profile extends out to 14\arcm\ from the
cluster center and shows no sign of a tidal  cutoff.  The inner profile is a
power-law with slope $-0.8$ steepening to $-1.7$ outside of 1\arcm. The mass
functions are for fields at 4\arcm\ and 11\arcm\ from the cluster center and
confirm the upturn in the mass function for stars less massive than about 0.4
M\solar.  There appears to be an excess of low-mass stars over higher-mass
stars in the outer field with respect to the inner, in qualitative agreement
with expectations for mass segregation.
\end{abstract}


\section{Introduction}
NGC~6397,  as the nearest globular cluster,  is a key observational target.
Deep and detailed observations can  be made of it with smaller investments of
time and  requirements for resolution than needed for more distant clusters.
Since the central surface brightness profile has a cusp, and
the estimated relaxation time is $\sim10^9 $yr, it is probable  that NGC~6397
has undergone significant dynamical evolution.  Further, its kinematics and
low metal abundance ([Fe/H] = -2.0) make this a definite member of the halo
population  of globular clusters. These factors have not been ignored and
several recent papers have turned their attention to this cluster.

Auri\`ere, Ortolani \& Lauzeral (1990) followed up on the work of
Auri\`ere (1982) in investigating the nature and distribution of the
stars in the core of NGC~6397. In addition to seeing the central cusp
(also confirmed by Djorgovski \& King 1986), they identified many of
the bright stars in the center as being blue stragglers. Lauzeral \et
(1992) found that it is these blue stragglers,  together with a
deficiency of red giants further out, which are responsible for a blue
color gradient  towards the center of the cluster (Auri\`ere 1982,
Djorgovski \et 1991). By removing these stars they were able to remove
the color gradient. The resulting surface brightness shows an apparent
core radius of 6\arcs.  Drukier (1993a) has shown  that the low number
of stars in the region of interest prevents the exclusion of a model
with an unresolved core.  Meylan \& Mayor (1991) published velocity
dispersions which they combined with existing surface density and
surface brightness profiles to constrain multi-mass King-Michie models
(Gunn \& Griffin 1979) of NGC~6397.  The high concentration of the
cluster makes this a difficult enterprise.  Anthony-Twarog, Twarog, \&
Suntzeff (1992) used CCD photometry in the Str\"omgren system to
explore the distance, age, reddening and metallicity of the cluster.

Da Costa (1982) produced star counts from photographic data to give a
surface density profile for the outer part of the cluster as well as a
luminosity function.  Fahlman \et (1989; hereafter FRST) continued this
work  with deep luminosity and mass functions from CCD observations.
Their mass function remains as the one probing to the lowest masses on
the main sequence and shows no sign of turning over to the limit of the
data at $0.12 M\solar$.  The mass function becomes much steeper for
stars less massive than $\sim 0.4M\solar$. In terms of the usual
power-law form
 \begin{equation}
 dN\propto m^{-(x+1)} dm,
 \end{equation}
where $x$ is the mass spectral index (MSI; $x=1.35$ for a Salpeter mass
function), $x$ increases for the stars less massive than $\sim
0.4M\solar$. Subsequent investigation of five more distant clusters
have shown similar upturns (Richer \et 1990 [M13,confirming Drukier \et
1988]; [M71], Richer \et 1991 [$\omega$ Cen, M5, and NGC 6752]).

It is to the earlier star count work that this paper serves as a follow
up.  We have undertaken an extensive set of observations of NGC~6397 in
the Cousins $I$--band in order to  produce both a surface density
profile (SDP)  and mass functions at additional radii. The new mass
functions allow us to confirm the increase in the mass spectral index
for low masses and to search for mass segregation.

Section 2 will discuss the observations and the general reduction
procedures used.  Section 3 discusses the resulting surface density
profile and section 4 the luminosity and mass functions. The final
section will briefly summarize the results of these star counts.

\section{Observations and reduction procedures}

The observations of NGC~6397 were obtained during an observing run at
the Las Campanas Observatory during May and June 1989. There were two
sets of observations, both in the Cousins $I$--band and using the TI
\#1 $800\times 800$ pixel CCD. The first data set consists of deep
observations of several 4.7 \sq\arcmin\ fields using the 2.5 m du Pont
Telescope. New observations of three cluster fields will be discussed
here. The second set of observations were obtained with the 1 m Swope
Telescope and covers  four overlapping 33.6 \sq\arcmin\ fields,
extending outwards from the cluster center, plus two regions further
out which overlap the 2.5 m fields. The positions and designations of the
fields are listed in Table~\ref{table3}.  Figure~\ref{map} shows the
locations of the program fields with respect to the cluster center.
The fields are labeled with their names from Table~\ref{table3} except
that ``du Pont'' has been shortened to ``duP''. All stars in the
program frames observed with $I<13$ have been marked by points. The
relative sizes of the points gives an indication of their magnitudes
and most of them can also be seen in Fig.~4 of Cannon (1974). In
addition to the program fields a field about $1\deg$ away from the
cluster and at the same galactic latitude was observed with both
telescopes. The star counts from this ``background field'' were used to
measure the contribution of non-cluster objects to the counts.
Observations of fields with standard stars were also take on nights
with photometric conditions. The 2.5 m observations were used to
construct luminosity and mass functions, while the purpose of the 1 m
data was to observe the surface density profile. On the 2.5 m, the
plate scale is $0\arcsper 162 {\rm\ pixel}^{-1}$. As the seeing was
usually around $1\arcs$ FWHM, this resulted in heavily over-sampled
star images. In order to improve the signal-to-noise ratio and speed up
processing, the frames were boxcar averaged $2\times 2$ following
debiasing and flat fielding, giving an effective plate scale of
$0\arcsper 324 {\rm\ pixel}^{-1}$.  For the 1 m frames the plate scale
is $0\arcsper 435 {\rm\ pixel}^{-1}$ and the seeing was typically
around $1\arcsper 7$ FWHM.

Debiasing was done using data from a second read of the CCD. Of this,
32 lines were read, averaged, and stored. For the TI CCD there is no
structure to this bias line, so the constant value was taken as the
bias level and subtracted. Flat fielding, image registration and
averaging, and image trimming were done using procedures within  IRAF.
The flat fields for most of the 2.5 m data were exposures of clouds
from the first night of the run. This flat field did not work very well
for the du Pont:bf field so a combination of this flat and dome flats
was used. Since the 2.5 m data were acquired under non-photometric
conditions, the exposures were weighted by their mean flux levels
during the averaging. The sigma clipping algorithm of the IRAF task
imcombine was used in order to remove cosmic rays. The deepest set of
exposures in each field were combined into two independent images for
that field and only stars found on both images were counted as
detections. While somewhat fainter stars could have been recovered if
all the data had been combined into a single frame, the advantage of
working with two images is that one can be more certain of the reality
of the recovered stars (Stetson 1991). In observing a mass function the
additional confidence conferred by this procedure far outweighs the
loss in limiting magnitude.

The data in fields Swope:nb, Swope:w and Swope:bf were taken under
photometric conditions and a colorless calibration was done using
standard stars in fields SA~110, SA~112, SA~113 and in the field of
PG~1323-085. The magnitudes used for the standards where preliminary
values of the ones in Landolt (1992). The differences between the
magnitudes used here and the published values are small and do not
affect the photometry.  Good results were obtained using the growth
curve program DAOGROW (Stetson 1990) and several other photometry
calibration programs kindly provided by P. B. Stetson. The
transformation to the standard system is good to within about $\pm
0.03$ mag.  The calibration for the corresponding 2.5 m frames was
transferred from the 1 m data. While the remaining four 1 m fields were
apparently observed under photometric conditions, it did not prove
possible to use the standards observed that night. Fortunately, the
Swope:sa and Swope:sb fields overlap that of FRST and a calibration,
good to within the $\pm 0.05$ quoted by FRST was obtained from
magnitudes of stars in that field. This calibration was also
transferred onto field du Pont:if.

Auri\`ere \et (1990) found the center of NGC~6397 to be within
1\arcs\ of their star 1. This star, identified with star 335 of Woolley
\et (1961),  was also taken as the center by Auri\`ere (1982).  Having
identified this star in field Swope:c, we adopted it as the cluster
center. The relative positions of all the other frames within the inner
region (ie. Swope:na, Swope:sa, Swope:sb, du Pont:if, and FRST) were
found with respect to Swope:c. The 2.5~m images were found to be
rotated very slightly with respect to the 1 m frames. The positions of
the Swope:nb and Swope:w fields with respect to the central four 1 m
fields were found using a photograph of NGC~6397. The photograph was a
print of a plate obtained at the prime focus of the 4m CTIO telescope
in the early seventies.  The positions of several stars in the two
pairs of regions were measured on the photograph with a measuring
engine and identified on the program frames. The required offset
between the origins of the Swope:sb and Swope:w fields, and between the
origins of the Swope:na and Swope:nb fields, were solved for in a least
squares fashion together with a constant scale factor.  No rotation was
apparent in the transformation relationship and in the final
calculation this was assumed to be zero. The resulting positions for
these fields with respect to the cluster accord well with a sketch of
the field positions made at the telescope. The positions of the du
Pont:n and du Pont:w fields, which are contained within the Swope:nb
and Swope:w fields respectively, were calculated with respect to the
positions of these 1 m fields.

The data were reduced using versions of  DAOPHOT and  ALLSTAR (Stetson
1987) in the usual manner. Two finding passes were made on the images
with a find threshold of $4\sigma$. Generally, a Moffat function
(Moffat 1969) with exponent 1.5 was used for the analytic portion of
the point spread function, plus a non-varying residual-look-up-table.
Occasionally, a linear variation in the look up table was used but it
did not materially affect the photometry. Each of the two independent
images for each field was reduced separately. Stars on the two lists
were matched and the mean offsets in magnitude and position were
calculated.  These offsets were then applied and a final star list for
each field was produced containing those stars found on both frames
with the centroids of the two images closer than 1 pixel.

In order to correct for incompleteness in the star counts, tests were
done using the usual procedure of adding artificial star images to the
program frames and then re-reducing them.  The artificial stars were
added with a magnitude distribution based on that estimated from the
initial reduction of the frame.  Ten percent of the number of stars
observed in the magnitude range of interest were added with random
magnitudes and positions to each artificial star frame. The same
artificial stars, with appropriate magnitude and position offsets, were
added to both images in a pair. Ten to thirty such pairs of frames were
reduced for each field.

The final star lists were prepared in the following manner. For each
field, regions surrounding saturated stars, diffraction spikes, charge
overflow columns and other defects, were identified. These areas have a
significantly higher rate of spurious detections. All artificial stars
added, and objects detected, in these regions were ignored during
further analysis. Also ignored were stars found with centers beyond the
edges of the frame since no artificial stars were added in these
areas.  The effects of these restrictions are small. The detection
lists for the two images in a pair were then shifted onto common
position and magnitude zero points. The stars on the lists were matched
and all objects found on both frames with position centroids within 1
pixel were deemed valid detections. The list of valid detections was
then compared with the list of positions and magnitudes of the
artificial stars added to that pair of frames. All valid detections
with positions within 1 pixel and magnitude differences less than 0.7
mag with respect to the artificial star list were counted as recoveries
of artificial stars. The remaining, presumably real, stars were put
into a separate list. Star counts were performed on the list of real
stars from each artificial star test separately and the results
averaged. Small variations in the number of stars were seen from test
to test due to variations in the local crowding conditions. Errors in
the star counts are a combination of the Poisson error and the standard
deviation in the number of stars recovered.

The observed luminosity function is a convolution of the true
luminosity function with the probability that a star of a given
magnitude will be found, and if so, with what observed magnitude. The
artificial star tests give an approximation to the probability function
and allow for the inversion of the observed luminosity function to give
the true one. The luminosity functions were corrected for
incompleteness using two techniques. In the simplest approach, all
stars added in a given magnitude bin and subsequently recovered were
counted as recoveries of stars with that magnitude. The ratio of the
number of recovered stars to the number added in that bin gives the
recovery rate, and the incompleteness correction factor is the inverse
of this. This method tends to underestimate the uncertainties in the
corrected counts and does not take into account the effects of stars
with true magnitudes lying in one bin being recovered in a neighboring
bin. This bin jumping can be corrected for in an approximate fashion
using techniques such as those described by Drukier \et (1988) and
Stetson \& Harris (1988).

The second method we used is a simplified version of the matrix
inversion method of Drukier \et (1988). As an intermediate method
between using the full matrix, as in Drukier {\it et\thinspace al.},
and compressing all the elements onto the diagonal, as in the approach
discussed above, a tri-diagonal recovery matrix was constructed. The
advantage of a tri-diagonal matrix over the full matrix is that it is
straightforward to invert it and to calculate the errors for the
inverted matrix. It is also  less susceptible to the  instability
associated with inverting a matrix containing many small elements. The
cost paid is that some information on the low-probability tails of the
measured-magnitude vs.  true-magnitude relations are lost, but not as
much as with the diagonal approach. The inverted matrix, together with
some assumptions about the recovery rates and magnitudes for stars in
the faintest bins, was used to calculate the completeness corrected
counts. The results of this method are consistent within the
uncertainties with those of the simpler approach  discussed above,
except that the error estimates and magnitude limit to the star counts
are more conservative using the matrix approach.  The final counts
presented in \S 4 were corrected using tri-diagonal matrices.

\section{Surface Density Profile}

The surface density profiles were produced from the 1 m data. Two
cutoff magnitudes were adopted and the resulting two surface density
profiles combined to give the final profile.  The first cutoff, $I=
14.0$, was adopted to obtain the surface density profile into the
center of the cluster.  Using the artificial star tests for the Swope:c
field, the completeness was calculated as a function of magnitude for
annuli about the center. In the central region, within a radius of 29
pixels (0\arcmper 21), the recovery rate was greater than 90\% to
$I=14.0$. A similar result was achieved for a pair of images of the
same field with exposure times of 5s. The second cutoff magnitude,
$I=15.5$, is discussed  below.

For fields Swope:na, Swope:sa, and Swope:sb we had longer exposure
images than are discussed here, but on these the brighter stars are
saturated and could not be counted in the luminosity function.
Therefore, the single short exposure images (averages of the two 5s
exposures obtained for each field) were used to produce the surface
density profiles. In the Swope:nb, Swope:w, and Swope:bf fields the
shallowest images, averages of two 30s exposures, were used. These
fields contained no stars bright enough to have poor photometry over
this length exposure.  For the Swope:c data the detected stars were
counted from the output of all 10 artificial star runs and an average
taken. For the bright stars, the variations due to crowding between the
final star lists on the artificial star frames was quite small,
confirming the incompleteness corrections.

The stars were counted in two sets of overlapping annuli to reduce the
effects of binning. The outer limits of the annuli were spaced
logarithmically with each annulus having a width of 0.2 dex. The radii
of the inner disks of the two sets of annuli were 0\arcmper 21 and
0\arcmper 17.   The areas of the intersections of the annuli with the
rectangular fields (we will refer to these as ``sections'') were
calculated and the star counts were converted to surface densities. The
adopted effective radius for each section is the radius which divides
the area of that section in half. This radius will vary between
sections on the same annulus, but from different fields, depending on
the geometry of the section. The field star density was measured on
Swope:bf. Twenty-three stars brighter than $I=14$ were found on that
frame giving a surface density of $0.69\pm 0.14$ stars per square arc
minute.

In order to account for areas in which no stars were seen, and to
eliminate the apparent anomalies associated with small number
statistics, any section of a field which contained fewer than 10 stars
was merged with its neighboring section. The area and effective radius
for the new, combined section were calculated, and the density and
error re-evaluated. The SDP after this merging is shown in
Fig.~\ref{i.lt.14} and listed in Table~\ref{table4}. The various
symbols indicate the frame of origin for each data point.

A set of surface density profiles were produced in the same way for a
set of centers offset by varying amounts from that adopted. Little
difference was seen for centers offset by up to $\pm 20$ pixels
(8\arcsper 7) in either or both directions. The outer SDP was little
affected by all these variations in center position and we conclude
that the errors in the determination of the relative positions of the
outer frames and the inner ones will have a negligible effect on the
surface densities.

When star counting is done on photographic plates, a reseau consisting
of sectored annuli is placed on the plate and the stars are counted
by sector. The density and uncertainty for a given annulus is
calculated from the counts for all the sectors in the annulus. Here,
with the exception of the central region, only a limited area of each
annulus was surveyed. In order to check the consistency of the star
counts and the assigned errors we split the lists of recovered stars
for Swope:c into quadrants about the cluster center and repeated the
counting procedure for each quadrant separately.  Figure~\ref{quad}
shows the results of this test. The symbols without error bars are the
densities for the four quarters of field Swope:c.  The line connects
the mean of the four measurements in each annulus and the error bars
are the standard deviations of the samples. The solid circles offset
+0.025 dex in radius are the overall densities from the full field
together with the uncertainties computed from Poisson statistics and
frame-to-frame variations in the counts. With the exception of the
innermost point, the quarter-to-quarter errors are within a factor of
two of the Poisson errors. There is excellent agreement in the errors
for mean points between $0\arcmper 25$ and $1\arcm$ but a larger
quadrant-to-quadrant variation is seen for the annuli beyond $1\arcm$.
The large scatter in the innermost point is an artifact of the linear
arrangement of the bright stars in the core of NGC~6397 (q.v. Auri\`ere
\et 1990).

The second surface density profile, containing all stars brighter than
$I=15.5$ and more than 20\arcs\ from the center, was produced in the
same manner as the $I\leq 14.0$ SDP. The inner radial limit is due to
incompleteness.  Beyond this point the counts from Swope:c are more
than 90\% complete between $I=15.0$ and $I=15.5$, and greater than 95\%
complete for all stars brighter that $I=15.5$. These incompletenesses
would change the final surface density profile by less than 0.05 dex,
well within the estimated errors. The magnitude cutoff of $I=15.5$ was
chosen since this is the magnitude of the turnoff on a roughly
calibrated $I-(V-I)$ color-magnitude diagram produced with data taken
at the same time as that of FRST. A field density of $2.06 \pm 0.25$
stars per square arc minute was derived from the Swope:bf data.  This
SDP is listed in Table~\ref{table5}.  The two surface density profiles
are shown in Fig.~\ref{sdp.all}.

The SDP of NGC~6397 seems to be best characterized by two power-laws
with a break at 1\arcm. Both single and multiple power-law models were
fit to the $I\leq 14.0$ SDP and for the multiple power-laws both the
cases of continuity and non-continuity across the break point were
tried. Quite consistently, the double power-law fits had significantly
lower values of reduced-$\chi^2$ (by a factor of two or more) than the
single power-laws. The preferred position for the break point in these
fits was at 1\arcm\ to within  10\%. The introduction of a third
power-law does not significantly improve the reduced-$\chi^2$.  For the
$I\leq 14.0$ SDP, the region with $r \leq 1\arcmper 26$ has a slope of
$-0.9\pm 0.1$ and the region $1\arcmper \leq r \leq 6\arcmper 3$ has a
slope of $-1.7\pm 0.1$.  The slopes are the mean of those calculated by
a weighted linear regression for the two sets of annuli independently.
The cutoff at $r=6\arcmper 3$ was chosen given that this is the radius
where the field and cluster densities are about equal and the
observational uncertainties in the SDP become large.  Over the region
$1\arcmper \leq r \leq 6\arcmper 3$ the $I\leq 15.5$ SDP has a slope of
$-1.73\pm 0.07$. The intercepts are better determined and give an
offset of $0.49\pm 0.04$ dex.

Since all the stars on both the $I\leq 14.0$ and $I\leq 15.5$ SDPs lie
at or above the turnoff they should have about the same mass  and so
should have the same radial density distribution. A comparison between
the two SDPs is shown in Fig.~\ref{sdp.all} where the offset of 0.49
dex has been applied to the $I\leq 14.0$ SDP. The background level for
the $I\leq 14.0$ SDP has been shifted by the same amount. The linear
fits to the profile are also shown in Fig.~\ref{sdp.all}.

Given the low galactic latitude of NGC~6397 ($b=-12\deg$),
determination of a limiting radius for this cluster is a difficult
proposition. For example, Da Costa (1979), from a single-mass King
model fit, finds a limiting radius of 38\arcm\ and Meylan \& Mayor
(1991) using multi-mass King models find a mean limiting radius of
95\arcm. These values are much larger than the limit of the new SDP
($14\arcm$). The present SDP steepens at the limit of the observations,
but does not extend nearly far enough to determine the limiting radius.
In view of the large uncertainties in the density and the preponderance
of field stars at these radii, much larger areas will need to be
observed in order to say anything about the limiting radius of NGC~6397.

\section{Luminosity and Mass Functions}

The luminosity functions (LFs) for fields du Pont:if, du Pont:n, du Pont:w,
and du Pont:bf were produced in accordance with the procedures outlined
in \S 2. The stars were counted in two sets of overlapping 0.5 mag
bins offset by 0.25 magnitudes in order to reduce the distortions due
to binning. For each field the true observed area, exclusive of the
ignored regions, was computed and the star counts adjusted to be the
number of stars which would be observed in a 4.67 \sq\arcmin\ field
($800\times 800$ pixels at a plate scale of 0\arcsper 162
pixel$^{-1}$).

At first glance the completeness corrected counts in
Tables~\ref{table6}--\ref{table9} may appear to present some
anomalies.  When the incompleteness corrections do not take into
account bin jumping, the corrected counts cannot be less than the raw
counts. The inclusion of the effects of bin jumping can lead to the
corrected number of stars in a bin being smaller than the original
number. To see that this can be the case consider the following. First,
for a given bin, bin $i$, the next fainter bin, bin $i+1$, generally
has more stars for a rising luminosity function as is usually the case.
Second, since the errors in magnitude increases with magnitude, a
higher proportion of the stars with true magnitudes in bin $i+1$ will
be observed with magnitudes in bin $i$ than the other way around. This
is so even if the distribution of magnitude differences are symmetric
about zero, but there is a tendency at the faintest magnitudes for
stars to be found too bright, since those sitting on positive noise
spikes have a better chance of being found than those sitting on
negative background fluctuations. The result of a rising luminosity
function and the increase of errors with magnitude is a net flux of
stars being found with magnitudes brighter than their true ones. When
the correction is made for bin jumping, bin $i$ could then end up with
fewer stars than it started with before the corrections.  Since some
fraction of observed stars are not recovered at all, the incompleteness
corrections (ie. the diagonal elements in the recovery matrix) will
increase the counts in bin $i$. However, if the fraction of the stars
observed in bin $i$ which belong in bin $i+1$ is greater than the
incompleteness correction, the number of stars in bin $i$ will decrease
after being corrected.

The background counts, converted to the number expected on a 4.67
\sq\arcmin\ field,  are tabulated in Table~\ref{table6}. In the bin
centered at  $I=22.5$, 50\% of the artificial stars were recovered with
magnitudes within 0.7 mag of their true values. For the next
independent 0.5 mag bin, centered at $I=23.0$, only 8\% of the
artificial stars were recovered. Since it is impossible, because of the
poor statistics, to estimate the number of stars in the $I=22.5$ bin
which originated in the $I=23.0$ bin, the $I=22.5$ bin could not be
used in the luminosity function. Similarly, for the overlapping bins
centered on the quarter magnitude, the last useful bin is at $I=21.75$.
Although the next bin has a recovery rate of 78\%, the one following
it, centered at $I=22.75$, has a recovery rate of only 26\%, and of
these, 44\% were found in the next higher bin, 12\% in the the next
fainter bin, and the remaining 44\% were found in the same magnitude
bin they were added in. The recovery rate drops quite quickly, so the
last usable bin usually has a rather high recovery fraction. As a
result of these considerations, the cutoff magnitude for the du Pont:bf
luminosity function is $I=22.25$; the last bin used being the one
centered at $I=22.0$.

Tables~\ref{table7}--\ref{table9} contain the raw and incompleteness
corrected star counts for the three program fields.The du Pont:if star
counts were found to have a 50\% recovery rate at $I=21.7$ and when the
bin jumping effects are taken into account the limiting magnitude of
the final bin is $I=21.5$. The 50\% recovery rate magnitude for the du
Pont:n data is $I=22.8$, and for the du Pont:w data it is $I=22.1$. The
magnitude limits of the incompleteness corrected star counts are 22.5
and 21.75 respectively.  The du Pont:n LF goes 0.75 mag deeper than the
du Pont:w LF because of better seeing and a longer total exposure.
Since the du Pont:w and du Pont:n fields are at about the same distance
from the cluster center, and the LFs for them have low signal-to-noise,
the LFs for these two fields have been averaged where they overlap and
a final, outer LF produced. The bins with centers with $I\geq 21.75$
are from the du Pont:n data only. The magnitude limit of the background
LF cuts off this combined LF, which will be referred to as the du
Pont:out LF, at $I=22.25$.

Figure~\ref{bkgd} shows a comparison of the star counts for the
background fields observed in this work and in FRST. The agreement is
good despite the two fields being on opposite sides of the cluster.
This agrees with the observation by Da Costa (1982) that the field
stars are evenly distributed in the vicinity of the cluster. The final,
background subtracted, cluster luminosity-functions are listed in
Table~\ref{table10} and are plotted in Figs.~\ref{if.lf} and
\ref{of.lf}. These two figures show the two new LFs and compares them
with the background star counts.  At the distance of the two outer
fields the number of field stars is equal to the number of cluster
stars, so deriving a more precise luminosity function will require
observing a much larger area.

Combined with the luminosity function of FRST, the new LFs give three
deep LFs for NGC~6397.  Figure~\ref{all.lf} brings together all three
of the observed deep LFs for NGC~6397. The two types of filled symbols
are the new LFs from this study, while the open symbols are the deeper
LF of FRST. No further normalizations have been applied; the offsets
between the LFs are a reflection of the overall decrease in density
with radius. For the magnitude range shown, all the stars should be on
the main sequence, since the turn-off lies at about $I=15.5$. The FRST
LF goes deeper because of the better seeing for that data, and because
of the more conservative approach to incompleteness corrections used
here.  The agreement between the three luminosity functions over their
common range indicates that the faint stars counted by FRST were real.
As discussed in \S 2, the conservative, tri-diagonal method for
incompleteness corrections gives substantially the same results as the
method used by FRST. Therefore, were the FRST star counts to be
recalculated with the more conservative method, the previous results
would stand.  The origin of the dip around $I=19$ in the du Pont:out LF
is unclear, but is seen, at about the same magnitude, in both fields.
Comparison with Fig.~\ref{bkgd} shows that it cannot be explained by
structure in the background star counts. The general depression in the
counts between $I\sim 18.5$ and $I\sim 20$ suggests that whatever is
causing the dip is not restricted to just the two low bins.

In order to convert the luminosity functions into mass functions a
distance modulus and a mass-luminosity law are required.  We adopt an
apparent $I$--band distance modulus of 12.0 as discussed in
Appendix~A.  The $I$--band mass-luminosity law of FRST, with an
extension to more massive stars taken from VandenBerg \& Bell (1985),
was used. This mass-luminosity relationship is for $Y=0.2$,
$Z=10^{-4}$, and an age of 16 Gyr.  Table~\ref{table11} lists the two
new mass functions. These are shown, together with the MF of FRST, in
Fig.~\ref{all.mf}.

{}From the more massive, upper main-sequence stars it is difficult to
see any sign of mass segregation, although there is some indication
that the outer field is more deficient in higher mass stars than the
two fields closer to the center. In relation to this, a series of
papers (Capaccioli, Ortolani, \& Piotto 1990, Piotto 1991, Capaccioli,
Piotto, \& Stiavelli 1993, Djorgovski, Piotto, \& Capaccioli 1993) has
been investigating correlations between globular cluster mass functions
and other cluster parameters such as galactic position and metallicity.
The mass function data are quantified in terms of the mass spectral
index over the range $0.5\leq M/M_{\sun} \leq 0.8$. Rather than the
observed mass spectral index at the particular location in the cluster
of the observation, global values are desired.  In the above series of
papers, the conversion from apparent to global MSI has been
accomplished following the prescription of Pryor, Smith \& McClure
(1986). In this technique multi-mass, King-Michie models are used to
calculate the apparent mass spectral index, $x_a$, as a function of
radius for a given  global mass spectral index, $x_g$, and
concentration.  There are several things to be concerned about with
this procedure. First, the mass spectrum used is very simple,
containing only five mass species, and will not necessarily be a good
match to the cluster mass function. Second, it only extends to
$0.2M\solar$ whereas FRST have counted stars to $0.12M\solar$ with no
indication that the mass spectrum ends. Third, different forms for the
global mass function will lead to different local-to-global corrections
and, while a power law may be a good approximation over a small mass
range, the observed mass functions show much more structure. Fourth, the
mass spectral index is computed over a very small mass range. Fifth,
in the case of clusters having central cusps, this technique is not
really practical since King models were never meant to represent
clusters undergoing core-collapse (King 1966).

For the cusp clusters Djorgovski \et (1993) use models with
concentration parameter $c=2.50$.   With the caveats discussed above in
mind we show in Fig.~\ref{psm} a Pryor \et (1986) style diagram for a
model with this concentration together with the mass spectral indices
over  $0.5\leq M/M_{\sun} \leq 0.8$ from the three mass functions in
Fig.~\ref{all.mf}. The core radius has been taken to be 6\arcs\ based
on the profile of Lauzeral \et (1992).  While a global value of
$x_g\sim 0$ is consistent with all three local measurements, the
agreement is very unsatisfactory.  There is a disturbing trend for the
``global'' value derived from this diagram to increase with radius.
Table~\ref{psm} lists the apparent mass spectral indices for the three
fields and their inferred global values from Fig.~\ref{psm} together
with their errors.  The mean is $\overline{x_g}=0.2\pm 0.6$. The
difficulty in finding a single, global, mass spectral index from the
three local values, indicates that the Pryor \et method is indeed not
useful in doing these conversions for, at least, post-core-collapse
clusters.

For the less massive stars, there is certainly an impression that the
outer field contains a relatively higher proportion of these than do
the two inner fields. For these stars there is a somewhat clearer
signal for mass segregation. Table~\ref{table12} gives the slopes of
weighted, least-square, power-law fits to the mass bins with masses
lower than 0.4, 0.32, and 0.28 $M\solar$, together with the numbers of
points fit and error estimates. The final point in the du Pont:out MF
has been excluded from this fitting due to its large uncertainty. The
errors for the slopes of the new Mfs are somewhat underestimated since
the points going into the fits are not independent.  For the du Pont:if
and FRST MFs the three slopes are quite consistent with one another,
and are also consistent with there being no mass segregation between
the two fields. As the high-mass cutoff is moved to lower masses the MF
from the outer fields gets steeper. The evidence for mass segregation
is strongest for the stars with $M<0.32 M\solar$ but is not highly
significant. There is also an indication in the three MFs that the mass
of the break in the MSI decreases with distance from the cluster
center. For the du Pont:if MF the upturn in the MF takes place at
around $0.4M\solar$, for the FRST MF this occurs between $0.4M\solar$
and $0.3M\solar$, and for the du Pont:out MF the data suggests an
upturn nearer to $0.3M\solar$. To some extent this is borne out by the
power laws fit to the MFs.  If this variation  of the break  with
radius is real, then this indicates that the break is not an artifact
of the knee in the mass-luminosity relation, as has been suggested
(Capaccioli \et 1993).

Drukier, Fahlman, \& Richer  (1992) introduced the concept of a
segregation measure, the ratio of  two mass functions, in order to
quantify the differences between mass functions independent of the
overall shape of the cluster's mass function. If the segregation
measure is computed between mass functions measured for two fields at
different radii, then the differences associated with mass segregation
will show up more clearly. If $N^a(m)$ is the number per unit mass of
stars with mass $m$ in field `$a$', then the radial segregation measure
for mass $m$ and field `a' with respect to field `b' is given by
 \begin{equation}
 S_r(m;a,b)\equiv\log[N^a(m)/N^b(m)].\label{segmes}
 \end{equation}
If $S_r(m_1;a,b)$ is larger than $S_r(m_2;a,b)$ then, with respect to
field `b', field `a' has an excess of stars of mass $m_1$  over stars
with mass $m_2$.  If, for example, field `a' lies further from the
cluster center than does field `b', then the segregation measure for
the low mass stars should be larger than that for the high mass stars
if mass  segregation has occurred in the cluster.  The segregation
measures between the three pairs of fields are shown in Fig.~\ref{sm}.
One difficulty with these diagrams are the large errors associated with
taking the ratio of two, already somewhat poorly determined,
quantities. The segregation measure between the inner two fields,
du Pont:if and FRST, is flat except for the most massive stars,
indicating that there is little mass segregation between the two
fields. The segregation measure between du Pont:if and du Pont:out
increases by 0.3 over a factor of 3.7 in mass.  We can define a
mass-normalized segregation measure over an observed mass range by
	\begin{equation}
	S_{r,m}(m_1,m_2;a,b)\equiv S_r(m_1;a,b)-S_r(m_2;a,b),
	\end{equation}
where $S_{r,m}(m_1,m_2;a,b)$ is the segregation measure of mass $m_1$
with respect to mass $m_2$ for field `a' with respect to field `b'. In
the event that a power law is a good fit to the mass function then, if
$x_a$ and $x_b$ are the MSIs in field `a' and `b' respectively,
	\begin{equation}
	x_a-x_b=-S_{r,m}(m_1,m_2;a,b)/\log(m_1/m_2).
	\end{equation}
If the full NGC~6397 mass functions could be characterized by single
power laws, then the mass-normalized segregation measure would imply
that the du Pont:out mass function has a MSI 0.5 larger than does the
du Pont:if mass function. The more complex structure observed in the mass
functions precludes such a inference, but the segregation measure does
support the conclusion that there is mass segregation  in NGC~6397.

\section{Summary}

We have observed 168.4 \sq\arcmin\ of NGC~6397 in the Cousins $I$--band
with deep photometry on a 13.7 \sq\arcmin\ sub-region.  The  shallower,
large-area observations have been used to construct a surface
brightness profile for the giants, subgiants and turn-off stars. The
limiting magnitude for these observations is $I=14$ for $r<0\arcmper
42$ and  $I=15.5$ for $0\arcmper 42 <r< 14\arcmper 1$.  The field star
surface density is equal to that of the cluster at about 7\arcm\ in
this magnitude range.

The core of the cluster is unresolved in these data. This is not
surprising  in that the innermost point is at 9\arcs\ and surface
brightness data suggest a core radius of order 6\arcs\ (Lauzeral \et
1992).  The surface density profile can be characterized by two power
laws with a break at about 1\arcm. Within this radius the power law
slope is $-0.9$. This slope is consistent with  those observed for
other globular clusters with central cusps (Djorgovski \& King 1986,
Lugger \et 1991). Beyond 1\arcm\ it is $-1.7$. There is no sign of a
tidal cut off to the limit of the observations at 14\arcm.

The deeper observations were used to produce local mass functions at
4\arcm\ and 11\arcm\ from the cluster center. Although these do not
extend to as low a mass as that of FRST---which was for a field at
6\arcmper 5 from the cluster center---they do confirm the change  in
slope of the mass function which occurs at about 0.4$M\solar$. No mass
segregation is seen between the 4\arcm\ and 6\arcmper 5 fields. There
is some indication of mass segregation between the 11\arcm\ field and
the others for the stars with $M<0.32M\solar$, but the difference in
the mass spectral indices is not significant. More stars will  need to
be counted in order to reduce the statistical uncertainties and confirm
the existence of any mass segregation in this cluster. For the stars
with masses in the range $0.5\leq M/M_{\sun} \leq 0.8$, the procedure
suggested by Pryor \et (1986) for converting apparent MSIs to their
global values does not give a consistent result. There is a trend for
the inferred global values to increase with radius.  This indicates
that this procedure should not be used for at least those clusters with
unresolved cores. Since, in the case of NGC~6397, neither the core nor
tidal radii are known, it is impossible to say what concentration model
to use.  In any case, King models were never designed for clusters with
such extreme concentrations (King 1966).

The prospects for further star count studies of NGC~6397 are
encouraging. As the discussion of the mass functions highlights, the
difficulty of CCD star counting has been the limited areas which can be
observed with telescopes large enough to get to faint magnitudes in
reasonable times.  With the advent of large area CCDs and mosaics of
CCDs this can be overcome and it will be possible to count enough stars
to make firm statements on the variation of NGC~6397's mass function
with radius and its surface density profile with magnitude.

Drukier \et (1992) examined a similar set of observations, in that case
for M71, in terms of a series of Fokker-Planck models.  They found that
they could not fit the M71 observations with any of the models they
considered since M71 showed too much mass segregation for a cluster
with such a large core radius.  NGC~6397 is in many ways a useful foil
to M71. Both lie at about the same galactocentric distance---7.4 kpc
for M71, 6.9 kpc for NGC~6397 (Webbink 1985)---but M71 has the
kinematics of a disk cluster while NGC~6397 belongs to the galactic
halo (Cudworth 1992). Further, NGC~6397 is the more massive and more
concentrated cluster of the two. A future paper (Drukier 1993b) will
compare  the observations discussed here with Fokker-Planck models and
comment further on both the current dynamical state of NGC~6397 and the
differences and similarities between it and M71.

\acknowledgments
This work was supported by grants from the Natural Sciences and
Engineering Research Council of Canada.

\newpage
\appendix
\section{$I$-band distance modulus to NGC 6397}

There are two sources for the distance modulus to NGC~6397. The first
is a venerable measurement of the magnitude of the horizontal branch by
Cannon (1974). Depending on the choice of horizontal branch
calibration, a distance modulus of $(m-M)_V = 12.3 \pm 0.3$ is obtained
(q.v. Harris 1980, Zinn 1985). More recently, Anthony-Twarog, Twarog,
\& Suntzeff (1992) derived an independent distance using Str\"omgren
photometry. Using main sequence fitting and the $M_V$, $(b-y)$
relationship method of Laird, Carney, \& Latham (1988) they found
$(m-M)_V = 12.1\pm 0.3$. With this distance modulus, the
color-magnitude diagram does not give a good fit to the theoretical
isochrones of VandenBerg \& Bell (1985) as modified for enhanced oxygen
abundance by McClure \et (1986). A better fit is achieved if $(m-M)_V =
12.4$ is used. Anthony-Twarog \et attribute some of the discrepancy to
the bolometric corrections used to transfer the luminosities onto the
observable plane. In any event, their distance moduli are consistent
with $(m-M)_V = 12.3$ adopted here.

NGC 6397  lies at $b=-11\degper 959$. Reddening measurements give
consistent values: Cannon (1974) found $E(B-V)= 0.18\pm 0.01$, van den
Bergh (1988) $E(B-V)=0.19\pm 0.02$, and VandenBerg, Bolte \& Stetson
(1990) $E(B-V)=0.19$.  VandenBerg \et also find evidence for a small
amount of variable reddening. From the data in their paper this appears
to be at the 0.013 magnitude level, and should not affect the results
here where a constant reddening of $E(B-V)=0.19$ is assumed. These
values for the distance modulus and reddening gives a heliocentric
distance of 2.2 kpc to NGC~6397. At this distance 1 pc subtends
$1\arcmper 6\pm 0\arcmper 2$, given the uncertainty in the distance
modulus.

Since the present observations are in the $I$--band, an apparent
distance modulus in that color is required. The formula in Dean,
Warren, \& Cousins (1978), extrapolated from O and B stars, implies
$E(V-I)/E(B-V)=1.35$ for the $(B-V)\sim 1.3$ typical for the lower main
sequence of NGC~6397 (Alcaino \et 1987).  Consideration of the entry
for K3~III stars  in Table~3 of Taylor (1986) gives a value of 1.38 for
this ratio. On the other hand, Grieve's (1983) calibration of the color
excess ratio for F and G supergiants gives a higher value, more like
1.6. This difference will only affect $(m-M)_I$ by 0.05 mag, well
within the uncertainties in the distance modulus itself. Given these
values for the ratio of the reddenings, we adopt an apparent $I$
distance modulus $(m-M)_I=12.0$. This is the same as the distance
modulus used by FRST.

\newpage
\begin{figure}
\caption[1]{Positions of the program fields observed for this study. The
coordinates are with respect to the adopted center marked by a `$+$'.
The dots mark the positions of all the stars observed with $I<13$ in
the program fields and have sizes related to their magnitudes. Most of
these also appear in the plate of Cannon (1974) and can be used to
locate these fields.  The abbreviation ``duP'' indicates the du Pont
fields listed in Table~\ref{table3}.  These were selected to avoid
bright stars.  The field marked ``FRST'' is that observed by Fahlman
\et (1989).}
\label{map}
\end{figure}

\begin{figure}
\caption[1]{
The surface density profile for all stars with $I<14$. The field star
density has been subtracted and is shown by the horizontal line. All
sections with fewer than ten stars (including those with no stars) have
been merged with their neighbors in the same field. As the relatively
empty sections are always at the edge of a field there is no ambiguity
in the direction of merging. Each density point is shown at the radius
which bisects the area of each (possibly composite) section and with
a symbol indicating the field of origin.}
\label{i.lt.14}
\end{figure}

\begin{figure}
\caption[1]{
A comparison of the central surface density profile for $I<14$ derived
from the four quadrants of the field Swope:c with that derived from the
entire field.  The various symbols without error bars are the surface
densities from star counts in the four quadrants. The line connects
their mean at each radius with error bars indicating the standard
deviation in the four measurements. The solid points, offset outward by
0.025 dex for clarity, are the surface densities from the whole field
with errors based on a combination of Poisson errors and frame to frame
errors as described in \S2.}
\label{quad}
\end{figure}

\begin{figure}
\caption[1]{
The combined surface density profiles for all stars with $I\leq 14.0$
(open circles) and $I\leq 15.5$ (filled circles). The former has been
shifted upward by 0.49 dex on the basis of the least-squares fits to
the slopes and intercepts over the region $0.0\leq\log r \leq 0.8$. The
field star surface densities to the same limiting magnitudes are also
shown.  The $I\leq 14.0$ field star density (horizontal dotted line)
has also been shifted by 0.49 dex. The overall fits to various radial
ranges of  the surface density profile are shown: solid line $\log
r\leq 0.1$, dot-dash line $0.0\leq\log r\leq 0.8$. The positions are
given in arc minutes. }
\label{sdp.all}
\end{figure}

\begin{figure}
\caption[1]{
Comparison of the field star counts from FRST (histogram) and
this study (field du Pont:bf; points). The two fields are at a distance
of 1\deg on either side of, and  at the same galactic latitude as the
cluster. Like the new luminosity functions, these star counts are
shown for two sets of overlapping half-magnitude bins.}
\label{bkgd}
\end{figure}

\begin{figure}
\caption[1]{
Luminosity function for the the inner field du Pont:if. The field star
counts are shown as a histogram for comparison. Two sets of overlapping
half-magnitude bins are shown.}
\label{if.lf}
\end{figure}

\begin{figure}
\caption[1]{
Mean luminosity function for the outer fields du Pont:n and du Pont:w.
The field star counts are also shown as a histogram. These fields are
about as far from the center of the cluster as is practical without
surveying very large areas.  Two sets of overlapping half-magnitude
bins are shown.}
\label{of.lf}
\end{figure}

\begin{figure}
\caption[1]{
All the NGC 6397 luminosity functions are compared. The three LFs are
from the inner field of this study, du Pont:if (filled circles);  that
of FRST at an intermediate radius (open circles); and  the mean of the
outer fields of this study (squares). For the two LFs from this study
two overlapping half-magnitude bins are shown. The offsets between the
luminosity functions are real and reflect the drop in stellar density
with radius. }
\label{all.lf}
\end{figure}

\begin{figure}
\caption[1]{
All the NGC 6397 mass functions are shown. The symbols and binning are
as in Fig.~\ref{all.lf}.}
\label{all.mf}
\end{figure}

\begin{figure}
\caption[1]{
Apparent mass spectral index vs. radius diagram following Pryor \et
(1986).  The five curves are labeled with the global values of the mass
spectral index for each model. A concentration of $c=2.5$ and core
radius of 6\arcs\ have been used.  The points are the observed values
of the MSI over the mass range $0.5\leq M/M_{\sun} \leq 0.8$. The
symbols for each mass function are as in Fig.~\ref{all.mf}.  While
$x_g\sim 0.$ is in marginal agreement with the observations, the
inferred global value for each point increases with radius.  This
inconsistency indicates that this is not a useful approach for doing
this  conversion.}
\label{psm}
\end{figure}

\begin{figure}
\caption[1]{
The segregation measures between the three pairs of fields are shown.
The radial segregation measure, $S_r(m;a,b)$, is defined by
eq.~(\ref{segmes}).  (a) $S_r(m;$  du~Pont:out,  du~Pont:if$)$.
(b) $S_r(m;$   du~Pont:out,   FRST$)$.
(c) $S_r(m;$   FRST,   du~Pont:if$)$. }
\label{sm}
\end{figure}

\clearpage
\begin{table}
\caption{Ignore this page, the tables are in tables.tex}
\label{table3}
\end{table}
\begin{table}
\caption{}
\label{table4}
\end{table}
\begin{table}
\caption{}
\label{table5}
\end{table}
\begin{table}
\caption{}
\label{table6}
\end{table}
\begin{table}
\caption{}
\label{table7}
\end{table}
\begin{table}
\caption{}
\label{table8}
\end{table}
\begin{table}
\caption{}
\label{table9}
\end{table}
\begin{table}
\caption{}
\label{table10}
\end{table}
\begin{table}
\caption{}
\label{table11}
\end{table}
\begin{table}
\caption{}
\label{pms}
\end{table}
\begin{table}
\caption{}
\label{table12}
\end{table}

\clearpage


\newpage
\begin{references}
\reference Alcaino, G., Buonanno, R., Caloi, V., Castellani, V., Corsi,
C. E., Iannicola, G., \& Liller, W. 1987, AJ, 94, 917
\reference Anthony-Twarog,  B.J.,  Twarog,  B.A., \& Suntzeff, N.B.
1992, AJ, 103, 1264
\reference Auri\`ere, M. 1982, A\&A, 109, 301
\reference Auri\`ere, M., Ortolani, S., \& Lauzeral, C. 1990, Nature,
344, 638
\reference Cannon, R. D. 1974, MNRAS, 167, 551
\reference Capaccioli, M., Ortolani, S., \& Piotto, G. 1991, A\&A, 244,
298
\reference Capaccioli, M., Piotto, G.P., \& Stiavelli, M. 1993, MNRAS,
261, 819
\reference Cudworth, K. M., 1992,  in of The Globular Cluster---Galaxy
Connection, ed. J. P. Brodie \& G. Smith (ASP Conf. Ser.), in press
\reference Da Costa, G. S. 1979, AJ, 84, 505
\reference \refrule 1982, AJ, 87, 990
\reference Dean, J. F., Warren, P. R., \& Cousins, A. W., 1978, MNRAS,
183, 569
\reference Djorgovski, S. \& King, I.R. 1986, ApJ, 305, L61
\reference Djorgovski, S., Piotto, G., \& Capaccioli, M. 1993, AJ, 105,
2148
\reference Djorgovski, S., Piotto, G., Phinney, E. S., \& Chernoff, D.F
1991 ApJ, 372, L41
\reference Drukier, G. A. 1993a, MNRAS, in press
\reference \refrule 1993b, in preparation
\reference Drukier, G. A., Fahlman, G. G., \& Richer, H. B. 1992, ApJ
386, 106
\reference Drukier, G. A., Fahlman, G. G., Richer, H. B., \&
VandenBerg, D. A. 1988, AJ, 95, 1415
\reference Fahlman, G. G., Richer, H. B., Searle, L., \& Thompson, I.
B. 1989, ApJ, 343, L49 (FRST)
\reference Grieve, G. 1983, Ph.D. thesis, University of Toronto
\reference Gunn, J. E. \& Griffin, R. F. 1979, AJ, 84, 752
\reference Harris, W. E. 1980, in IAU Symp. 85, Star Clusters, ed. J.
E. Hesser (Dordrecht: Reidel), 81
\reference King, I.  1966, AJ, 71, 64
\reference Laird, J. B., Carney, B. W., Latham, D. W. 1988, AJ, 95, 1843
\reference Landolt, A.U. 1992, AJ, 104, 350
\reference Lauzeral, C., Ortolani, S., Auri\`ere, M., Melnick, J. 1992,
A\&A, 262, 63
\reference Lugger, P.M., Cohn, H.N., Grindlay, J.E., Bailyn, C.D. \&
Hertz, P.L.1991, in The Formation and Evolution of Star Clusters, ed.
K. Janes (ASP Conf. Ser., 13), 414
\reference McClure, R. D., VandenBerg, D. A., Smith, G. H., Fahlman, G.
G., Richer, H. B., Hesser, J. E., Harris, W. E., Stetson, P. B., \&
Bell, R. A. 1986, ApJ, 307, L49
\reference Meylan, G., \& Mayor, M. 1991, A\&A, 250, 113
\reference Moffat, A. F. J. 1969, A\&A, 3, 455
\reference Piotto, G.P. 1991, in The Formation and Evolution of Star
Clusters, ed. K. Janes (ASP Conf. Ser., 13), 200
\reference Pryor, C., Smith, G.H. \& McClure, R.D. 1986, AJ, 92, 1358
\reference Richer, H. B., Fahlman, G. G., Buonanno, R., \& Fusi Pecci,
F. 1990, ApJ, 359, L11
\reference Richer, H. B., Fahlman, G. G., Buonanno, R., Fusi Pecci, F.,
Searle, L., \& Thompson, I. B., 1991, ApJ, 381, 147
\reference Stetson, P. B. 1987, PASP, 99, 191
\reference \refrule 1990, PASP, 102, 932
\reference \refrule 1991, in The Formation and Evolution of Star
Clusters, ed. K. Janes (ASP Conf. Ser., 13), 88
\reference Stetson, P. B. \& Harris, 1988, AJ, 96, 909
\reference Taylor, B. J. 1986, ApJS, 60, 577
\reference van den Bergh, S. 1988, AJ, 95, 106
\reference VandenBerg, D. A. \& Bell, R. A. 1985, ApJS, 58, 561
\reference VandenBerg, D. A., Bolte, M., \& Stetson, P. B. 1990, AJ,
100, 445
\reference Webbink, R. F. 1985, in IAU Symp. 113, The Dynamics of Star
Clusters, ed. J. Goodman \& P. Hut (Dordrecht: Reidel), 541
\reference Woolley, R., Alexander, J. B., Mather, L., \& Epps, E. 1961,
Roy. Obs. Bull., 43, E303
\reference Zinn, R. 1985, ApJ, 293, 424
\end{references}
\end{document}